%% file: main.tex
\documentclass[11pt,a4paper]{article}
\usepackage{amsmath,amssymb,amsbsy,bm,upgreek,nicefrac,fullpage}
\usepackage{todonotes,microtype}
\usepackage{comment}
\usepackage{authblk}
\usepackage{lipsum}  
\usepackage{blkarray}
\usepackage{tikz}
\usepackage[utf8]{inputenc}
\usepackage[a4paper,margin=2cm]{geometry}
\usepackage{graphicx}
\usepackage{float}
\usetikzlibrary{trees,positioning,shapes,shadows,arrows}
\usetikzlibrary{fit}
\usepackage{url} 

\usepackage{hyperref}
\usepackage[capitalise]{cleveref}
\crefrangeformat{equation}{Eqs.~#3#1#4 --~#5#2#6}
\graphicspath{{./Figures/}}

\newcommand{\ZZ}{\mathbb Z}
\newcommand{\RR}{\mathbb R}

\title{Music Composition Using Quantum Annealing}

\author[1]{Ashish Arya}
\author[2,3]{Ludmila Botelho}
\author[4]{Fabiola Ca\~nete}
\author[5]{Dhruvi Kapadia}
\author[2,6]{\"Ozlem Salehi \thanks{osalehi@iitis.pl} }	

\affil[1]{ICT Academy, Indian Institute of Technology, Kanpur, India}
\affil[2]{Institute of Theoretical and Applied Informatics\\ 
	Polish Academy of Sciences\\ 
 Gliwice, Poland}
\affil[3]{Joint Doctoral School, Silesian University of Technology, Gliwice, Poland}
\affil[4]{Benemérita Universidad Autónoma de Puebla. Puebla City, Mexico}
\affil[5]{Sarvajanik College of Engineering and Technology, India}
\affil[6]{QWorld Association, Tallinn, Estonia }
\date{}

\begin{document}

\maketitle

 \begin{abstract}
With the emergence of quantum computers, a new field of algorithmic music composition has been initiated. The vast majority of previous work focuses on music generation using gate-based quantum computers. An alternative model of computation is adiabatic quantum computing (AQC), and a heuristic algorithm known as quantum annealing running in the framework of AQC is a promising method for solving optimization problems. In this chapter, we lay the groundwork of music composition using quantum annealing. We approach the process of music composition as an optimization problem. We describe the fundamental methodologies needed for generating different aspects of music including melody, rhythm, and harmony. The discussed techniques are illustrated through examples to ease the understanding. The music pieces generated using D-Wave quantum annealers are among the first examples of their kind and presented within the scope of the chapter. The text is an unedited pre-publication version of a chapter which will appear in the book "Quantum Computer Music", Miranda, E. R. (Editor).
 \end{abstract}

\input{introduction}

\section{Background} \label{sec: back}
\input{optimization}

\input{quantum_annealing}

\input{markov}

\section{Music Composition as an Optimization Problem} \label{sec: opt}

\input{music_optimization}

\section{Music Composition Using Quantum Annealing} \label{sec: gen}
\input{music_generation}

\section{Conclusion and Future Work} \label{sec: conc}
\input{discussion}

\section*{Acknowledgements} 
\input{acknowledgement}


\bibliographystyle{ieeetr}
\bibliography{music}

\end{document}

%% file: introduction.tex
\section{Introduction}

Music composition can be thought of, in a very simplistic manner, as a creative process where one puts sounds and silences together that results in a sequence that is aesthetic or pleasant to the ear. Over the years, it has been discovered that music pieces that are soothing and pleasing follow some rules and possess common patterns. Those rules have evolved and solidified up to some extent over the centuries. Yet, there is still certain flexibility keeping open room for creativity.

Being able to identify some rules and common patterns to guide the music composition process is one of the keystones of the field of algorithmic music composition. The seeds of computer-generated music were sown at the end of the nineteenth century by Ada Lovelace, the first computer programmer, who put forward the idea that Babbage’s prototype computer, the analytical engine, ``might compose elaborate and scientific pieces of music of any degree of complexity or extent'' \cite{menabrea1842sketch}. Yet, this dream was not realized until the 1950s, when the first computer-generated music piece, the Iliac Suite, was composed \cite{hiller1979experimental}. Following the advancements in computer science, various methodologies have been explored for generating music, including stochastic approaches, rule-based systems, evolutionary algorithms, and machine learning \cite{papadopoulos1999ai}. The emergence of quantum computers heralds a new addition to this sequel.

With quantum computers being an alternative tool for generating music, we are witnessing the growth of a new field referred to as quantum computer music \cite{miranda2020quantum}. Although the term comprises using quantum computers to generate, perform, and listen to music, we will focus on music generation harnessing the quantum computing paradigm in the scope of this chapter.  
 
Gate-based computing and adiabatic quantum computing are the two popular computational models in quantum computing. In the gate-based model, states of the qubits are manipulated through unitary operations, the so-called quantum gates. Most of the work undertaken so far is based on gate-based quantum computing. In \cite{kirke2018programming}, a simple algorithm named GATEMEL is developed to generate music using IBM quantum computers.  A classical-quantum algorithm is introduced in \cite{kirke2019applying}, which uses Grover's search and follows a rules-based approach for composing music. The recent work by Miranda and Basak uses quantum walks \cite{miranda2021quantum} and another novel approach, quantum natural language processing, is used in \cite{miranda2021quantum2}. 

Adiabatic quantum computing (AQC) is an alternative to the gate-based model of computation. In AQC,  the computation is driven by applying an external magnetic field. We will be particularly interested in Quantum Annealing \cite{aharonov2008adiabatic,doi:10.1126/science.1057726}, a heuristic algorithm based on the AQC model for solving optimization problems. Using quantum annealing for creating music requires the formulation of the music creation process as an optimization problem. It has been previously used to generate harmony by Kirke and Miranda  \cite{kirke2017experiments}.

In this chapter, we develop new methodologies for generating music using quantum annealing. We present the main building blocks for formulating rules-based music generation as an optimization task. We consider the problem from various aspects, including the composition of melody, rhythm, and harmony. Using D-Wave quantum annealers, we generate music pieces that are displayed in the course of the text. The presented material contributes to the field as the first text focusing on the role of Quantum Annealing for music composition in an extensive manner. 

The rest of the chapter is structured as follows. We begin with a background on optimization problems, Quadratic Unconstrained Binary Optimization and Integer Linear Programming formulations, Quantum Annealing, and Markov Random Fields in Sec. \ref{sec: back}. In Sec. \ref{sec: opt}, we present a review of the classical approaches for music composition in the framework of optimization. We describe the fundamental techniques and formulations for music composition in Sec. \ref{sec: gen}. Finally, we conclude with a discussion and suggestions on future work in Sec. \ref{sec: conc}.

%% file: optimization.tex
In this section, we present the necessary background on optimization and quantum annealing. Before
we move on to the discussion, let us get familiar with some notations that will be useful in the rest of
the chapter.

By $\ZZ$ and $\RR$, we denote the set of integers and real numbers respectively. Given that $x$ is a column vector defined over $\ZZ$ or $\RR$, $x_i$ denotes the $i$'th element of $x$ and $x^\top$ denotes the transpose of $x$. When $A$ is a matrix, $A_{ij}$ denotes the element in the $i$'th row and $j$'th column of $A$. By $[n]$, we denote the set of integers $\{1,2,\dots,n\}$.

\subsection{Combinatorial Optimization}\label{sec:optimization}
Combinatorial optimization aims to minimize or maximize an objective function that is defined over a discrete set. They appear in almost every field of science and engineering, including logistics, supply chain management, transportation, and finance. Many of the well-known problems like the Knapsack Problem, Travelling Salesperson Problem, and Graph Coloring are optimization problems, and they have applications in scheduling, resource allocation, assignment, and planning. All of the mentioned problems are NP-Hard and become easily intractable. A vast number of optimization techniques were performed on classical computers to solve optimization problems, including both exact and heuristic methods. Quantum computing is offering novel approaches for solving optimization problems, as we will discuss in the following subsections.  

\subsubsection{Integer Linear Programming}\label{sec: ilp}

Most of the time when we optimize a function, the optimization is subject to some constraints. \textit{Integer Linear Programming} (ILP) is a mathematical model for problems defined over integer variables with a linear objective function and a set of linear constraints. Formally, a linear program is defined as
\begin{alignat*}{3}
&\text{minimize} \hspace{1em}&& \sum_{j} c_j y_j \\
&\text{subject to} \hspace{1em }&& \sum_{j} A_{ij}y_j \leq b_i, \hspace{1em} i=1,\dots,m \\
&{} &&y_j \geq 0, y_j \in \ZZ   
\end{alignat*}
where $A_{ij}\in \RR$, $b_i\in \RR$, $ c_j \in \RR $. ILP problem is known to be NP-Hard. In case the objective function is a quadratic polynomial, then the model is named as Integer Quadratic Programming (IQP). There are exact algorithms, including cutting plane and branch-and-bound methods, and heuristic algorithms like Simulated Annealing and Ant Colony Optimization for solving ILPs.

\subsubsection{Quadratic Unconstrained Binary Optimization}

Quadratic Unconstrained Binary Optimization (QUBO) problems involve an objective function defined over binary variables consisting of linear and quadratic terms. As opposed to ILP, there are no constraints. QUBO formulation has become extremely popular with the advent of quantum annealing, as will be explained in the upcoming subsections. An extensive list of problems and their QUBO formulations are presented in \cite{lucas2014ising}.

Formally QUBO formulation is defined as
\begin{align*}
f(x) =  \sum_{i\leq j} x_i Q_{ij} x_j,
\end{align*}
where  $x_i \in \{ 0,1 \}$  and $ Q $ is a real square upper triangular matrix. The above sum can be equivalently expressed as $x^\top Qx$ where $x$ is the vector of binary variables. The goal is to find the binary vector $x$ that minimizes $f(x)$.



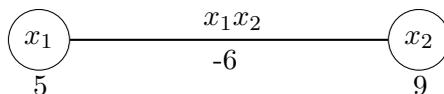
\begin{figure}[H]
	\centering
	\begin{tikzpicture}
	\node[circle,draw] (a) at (1,1) {$x_1$};  
	\node[circle,draw] (b) at (6,1){$x_2$};
	\draw[thick, black](a) -- (b);
	\coordinate [label=below:5] (a) at (1.02,0.65);
	\coordinate [label=below:-6] (a) at (3.45, 1.00);
	\coordinate [label=below:9] (a) at (6.02, 0.65);
	\coordinate [label=above:$x_1x_2$] (a) at (3.55, 1.00);
	\end{tikzpicture}
	\caption{A visualization for the function $f(x_1,x_2)$, where the circles represent the variables, the numbers below the circles are the coefficients of the linear terms, and the label of the edge between is the quadratic coefficient.}
	\label{fig:obj}
\end{figure}

Let us look at an example involving two variables. Suppose that we aim to minimize the given objective function
\begin{equation}
f(x_1,x_2) = 5x_1 + 9x_2 -6x_1x_2.
\end{equation}
A visual representation of $f(x_1,x_2)$ is given in Fig.~\ref{fig:obj}.

Let us identify matrix $Q$ for the given $f(x_1,x_2)$. Here, $5x_1$ and $6x_2$ are the linear terms. Note that $5x_1 = 5x_1^2$ and $9x_2 = 9x_2^2$ since $x_1$ and $x_2$ are binary variables. We will place 5 and 6 to the diagonals of the $Q$ matrix and the 1'st row and 2'nd column of the matrix is $-6$ due to the term $-6x_1x_2$ as shown below:
\begin{equation}
    	\begin{blockarray}{ccc}
    x_1 & x_2 \\
    \begin{block}{(cc)c}
     5  & -6 & x_1 \\
     0  & 9 & x_2 \\
    \end{block}
    \end{blockarray}.
\end{equation}
Observe that $x^\top Qx$ yields us $f(x_1,x_2)$:

$$
\begin{bmatrix} 
	x_1 ~~
	x_2 
	\end{bmatrix}
	\begin{bmatrix} 
	5 & -6 \\
	0 &  9 \\
	\end{bmatrix}
	\begin{bmatrix} 
	x_1 \\
	x_2 \\
	\end{bmatrix}
	= \begin{bmatrix} 
	x_1 ~~
	x_2 
	\end{bmatrix}
	\begin{bmatrix} 
	5x_1-6x_2 \\
	9x_2 \\
	\end{bmatrix}
     =
	5x_1^2-6x_1x_2 + 9x_2^2 
$$
Let us find out the values for $x_1$ and $x_2$ that minimizes $f(x_1,x_2)$ analytically. There are four possible cases:

$x_1 = 0$ and $x_2 = 0$;
\begin{equation}
f(0,0) = 5\cdot 0 + 9\cdot 0 - 6\cdot 0\cdot 0 = 0
\end{equation}

$x_1 = 0$ and $x_2 = 1$;
\begin{equation}
f(0,1) = 5\cdot 0 + 9\cdot 1 - 6\cdot 0\cdot 1 = 9
\end{equation}

$x_1 = 1$ and $x_2 = 0$;
\begin{equation}
f(1,0) = 5\cdot 1 + 9\cdot 0 - 6\cdot 1\cdot 0 = 5
\end{equation}

$x_1 = 1$ and $x_2 = 1$;
\begin{equation}
f(1,1) = 5\cdot 1 + 9\cdot 1 - 6\cdot 1\cdot 1 = 8
\end{equation}

The case $x_1=0$ and $x_2=0$ minimizes the objective function and yields the lowest possible value. 

\subsubsection{Converting ILP to QUBO}\label{sec: ilptoqubo}

Any ILP or IQP can be converted into QUBO. This conversion is helpful as ILP formulations for many problems are already known, and it is often easier to express optimization problems through constraints. Now, let us explain how ILP and IQP can be processed as a QUBO. The necessary steps are described
on the diagram in Fig. \ref{fig:diagram}.

\usetikzlibrary{shapes.geometric, arrows}
\tikzset{
	basic/.style  = {draw, text width=2cm, drop shadow, font=\sffamily, rectangle}}

\tikzstyle{V} = [rectangle, rounded corners, minimum width=3cm, minimum height=1cm,text centered, draw=black, fill=white!30]
\tikzstyle{W} = [rectangle, rounded corners, minimum width=3cm, minimum height=1cm,text centered, draw=black, fill=white!30]
\tikzstyle{X} = [rectangle, rounded corners, minimum width=3cm, minimum height=1cm,text centered, draw=black, fill=white!30]
\tikzstyle{Y} = [rectangle, rounded corners, minimum width=3cm, minimum height=1cm,text centered, draw=black, fill=white!30]
\tikzstyle{Z} = [rectangle, rounded corners, minimum width=3cm, minimum height=1cm,text centered, draw=black, fill=white!30]
\tikzstyle{arrow} = [thick,->,>=stealth]
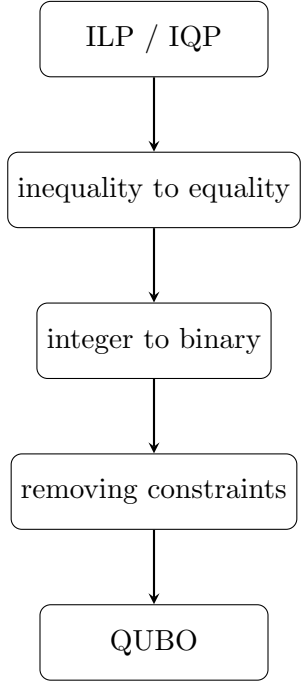
\begin{figure}[H]
	\centering
	\begin{tikzpicture}[node distance=2cm]
	\node (a) [V] {ILP / IQP};
	\node (b) [W, below of=a] {inequality to equality};
	\node (c) [X, below of=b] {integer to binary};
	\node (d) [Y, below of=c] {removing constraints};
	\node (e) [Z, below of=d] {QUBO};
	\draw [arrow] (a) -- (b);
	\draw [arrow] (b) -- (c);
	\draw [arrow] (c) -- (d);
	\draw [arrow] (d) -- (e);
	\end{tikzpicture}
	\caption{Steps of the conversion from ILP/IQP to QUBO.}
	\label{fig:diagram}
\end{figure}

As we will see, there are three steps of this process. Let us discuss each step briefly.
\begin{enumerate}
\item \textbf{Inequality to Equality:} The first step of the conversion is turning inequality constraints into equality constraints by adding so-called \textit{slack} variables. Slack variables compensate the difference between the left-hand side and the right-hand side of the inequality if the inequality is satisfied. Suppose that we have the linear inequality constraint: 
\begin{equation}
\sum_{i=1}^ka_i y_i \leq b.
\end{equation}
Then by adding slack variable $\xi$, we obtain $\sum_{i=1}^ka_i y_i + \xi = b$. Note that when   $\sum_{i=1}^ka_i y_i=b$, then $\xi$ is 0, which is the lower bound for $\xi$. The upper bound should be large enough so that the equality is satisfied when $\sum_{i=1}^ka_i y_i$ has the lowest possible. Further details can be found in \cite{salehi2021unconstrained}.
\item \textbf{Integer to Binary:} This step involves expressing each bounded integer variable using a set of binary variables \cite{karimi2019practical}. Suppose $y$ is an integer variable with the lower bound $\underline y$ and upper bound $\overline y$. Note that $y$ is equivalent to the following term:
\begin{equation}\label{eq:yterm}
\underline y+\sum_{i=0}^{k-2}2^i x_{i}+\bigl (\overline{y}-\sum_{i=0}^{k-2}2^i\bigr)x_{k-1},
\end{equation}
where $k = \lceil \log_2(\overline y-\underline y+1)\rceil $, and $x_{i}$ are the newly introduced binary variables. Hence, we transform our problem into an equivalent one expressed using binary variables.

\item \textbf{Removing Constraints} Converting constrained programs into unconstrained ones dates to 1970's \cite{geoffrion2010lagrangian}. To create a constraint-free formulation, the penalty method is used. The idea is to introduce penalties to the objective function when the constraints are violated. In Table \ref{table:penalties}, a list of linear inequality/equality constraints and how they are incorporated into the objective function is given in Table \ref{table:penalties}.

\begin{table}
\begin{center}
\begin{tabular}{ |c|c| } 
 \hline
Constraint & Equivalent Penalty  \\ 
 \hline
 $x_1+x_2 \leq 1$ & $x_1x_2$ \\
 \hline
 $x_1+x_2 \geq 1$ & $1-x_1-x_2+x_1x_2$ \\
 \hline
 $x_1+x_2 = 1$ & $1-x_1-x_2+2x_1x_2$ \\
 \hline
 $x_1 \leq x_2$ & $x_1-x_1x_2$ \\
 \hline
 $x_1 + x_2 + x_3 \leq 1$ & $x_1x_2 + x_2x_3 + x_3x_1$ \\
 \hline
 $x_1 = x_2$ & $x_1+x_2-2x_1x_2$ \\
 \hline
\end{tabular}
\caption{Some linear constraints and equivalent penalties \cite{glover2019quantum}.}
\label{table:penalties}
\end{center}
\end{table}

For instance, let us say the objective function is $f(x_1,x_2)=(x_1+x_2)^2$ and we have the constraint $x_1+x_2 \leq 1$ where $x_1$ and $x_2$ are binary variables. The equivalent QUBO formulation is given by $f(x) + Px_1x_2$ where $P$ is th non-negative, real, sufficiently large  penalty value. Note that the constraint enforces that both $x_1$ and $x_2$ are not equal to 1 at the same time. Whenever this is the case, a penalty of $P$ is added to the objective in the QUBO formulation, due to term $Px_1x_2.$

Now let us see how to deal with general linear constraints of the general form. As we have seen already a method to convert inequalities into equalities, it is enough to consider linear equality constraints only. Suppose we have the following minimization problem:
\begin{alignat*}{3}
&\text{minimize} \hspace{1em}&&  f(x) \\
&\text{subject to} \hspace{1em }&& g_i(x) = c_i , i \in [m] \\
&{} &&x_i \in \ZZ   
\end{alignat*}
where $f(x)$ is a linear or quadratic objective function, $g_i(x)=c_i, i \in [m]$ is a set of linear equality constraints. The equivalent unconstrained formulation is given as
\begin{equation}
    \text{minimize} \hspace{1em} f(x) + \sum_{i\in [m]}P_i(g(x)-c_i)^2
\end{equation}
where $P_i$'s are the associated penalty parameters. The larger the penalty values is, the larger the increase in the energy will be in the case each constraint is violated. If $P_i$ is not large enough, then it is likely that the corresponding constraint is violated in favor of the objective function.
\end{enumerate}

\subsubsection{Quadratization}
Sometimes, one needs to deal with formulations consisting of not only quadratic but also higher-order terms. Such unconstrained formulations defined over binary variables are called Higher-Order Binary Optimization (HOBO) problems. By introducing auxiliary variables 
$y_1$,$y_2$,..,$y_m$, it is possible to reduce the problem of minimizing HOBO into the problem of minimizing QUBO. There are several proposed methods for quadratization. We will explain the one by Rosenberg et al. \cite{rosenberg1975reduction}. 

1. Find two variables $x_i$ and $x_j$, such that $x_ix_j$  appears in a term with the degree at least 3.

2. Replace all terms that contain $x_ix_j$, with $y_{ij}$ which belongs to \{0,1\}.

3. Add the penalty term $P(x_ix_j-2x_iy{ij} - 2x_jy_{ij} + 3y_{ij})$, where $P$ is the penalty constant.

%% file: quantum_annealing.tex
\subsection{Quantum Annealing}

Let us now explain the theoretical background behind Quantum Annealing, and also give insights on solving optimization problems using the publicly available quantum annealers.
\subsubsection{Motivation}
Many interesting questions can be boiled down into optimization problems. Let us consider the Travelling Salesperson Problem, where the aim is to find the route with minimum distance (cost) that passes through each city exactly once and returns to the first city, given a set of cities and the distance between them. Imagine that the salesperson has to visit 50 different cities and return to the starting point. There are $50!$ different routes that can be taken by the salesperson, and finding the route with the smallest cost by calculating all the possibilities is a costly method in terms of time and energy; and for many complex problems, it is almost impossible.\footnote{Although there are more clever methods than trying all the routes one by one, the best known exact algorithm has still exponential time complexity.} If we consider the $50!$ possible routes as our search space and call the cost associated with each route the energy, the problem can be framed as an energy minimization problem. Hence, finding the answer is equivalent to looking around the hills and valleys in an energy landscape for the lowest point.

\emph{Quantum Annealing} (QA) is a promising heuristic algorithm for solving optimization problems by taking advantage of properties peculiar to quantum physics like quantum tunneling, entanglement, and superposition \cite{apolloni1989quantum, kadowaki1998quantum}. This approach for finding the minimum of an objective function is based on the principles of Adiabatic Quantum Computing (AQC). AQC is an alternative to the gate-based computation, and the two computational models are equivalent in terms of power \cite{aharonov2008adiabatic,doi:10.1126/science.1057726}. Unlike the gate-based model in which the evolution of the system is discretized by the application of gates, AQC is a continuous-time process, and it has an analog nature. The computation is driven by the application of a time-dependent external magnetic field. In QA, the evolution takes a restricted form of AQC, and some conditions of AQC, such as the computation taking place in a totally closed system, are relaxed.

Quantum Annealing is the quantum counterpart of Simulated Annealing (SA) \cite{kirkpatrick1983optimization}, a probabilistic technique for optimization. In SA, the exploration of the search space is governed by a temperature parameter that is slowly decreased throughout the process. In the lower temperatures, transitions between states occur less frequently. The term ``annealing'' originates from a metallurgy technique, where a material is heated and then cooled in a controlled manner to alter its physical properties, analogous to the temperature parameter in SA. SA harnesses thermal fluctuations to climb over the valleys and escape from local minima. In quantum annealing, this process of moving towards the low energy states is achieved by quantum tunneling. In Fig. \ref{fig:qa}, how the both algorithms move through the energy landscape is depicted.

\tikzset {_etbxlmyj4/.code = {\pgfsetadditionalshadetransform{ \pgftransformshift{\pgfpoint{89.1 bp } { -108.9 bp }  }  \pgftransformscale{1.32 }  }}}
\pgfdeclareradialshading{_345n2kpid}{\pgfpoint{-72bp}{88bp}}{rgb(0bp)=(1,1,1);
	rgb(0bp)=(1,1,1);
	rgb(25bp)=(0,0,0);
	rgb(400bp)=(0,0,0)}


\tikzset {_m5kxfk23u/.code = {\pgfsetadditionalshadetransform{ \pgftransformshift{\pgfpoint{89.1 bp } { -108.9 bp }  }  \pgftransformscale{1.32 }  }}}
\pgfdeclareradialshading{_4bkrn46oe}{\pgfpoint{-72bp}{88bp}}{rgb(0bp)=(1,1,1);
	rgb(0bp)=(1,1,1);
	rgb(25bp)=(0,0,0);
	rgb(400bp)=(0,0,0)}


\tikzset {_5stw8zhpc/.code = {\pgfsetadditionalshadetransform{ \pgftransformshift{\pgfpoint{89.1 bp } { -108.9 bp }  }  \pgftransformscale{1.32 }  }}}
\pgfdeclareradialshading{_t493ccfgj}{\pgfpoint{-72bp}{88bp}}{rgb(0bp)=(1,1,1);
	rgb(0bp)=(1,1,1);
	rgb(25bp)=(0,0,0);
	rgb(400bp)=(0,0,0)}


\tikzset {_4cx8zkr77/.code = {\pgfsetadditionalshadetransform{ \pgftransformshift{\pgfpoint{89.1 bp } { -108.9 bp }  }  \pgftransformscale{1.32 }  }}}
\pgfdeclareradialshading{_31uveu6mu}{\pgfpoint{-72bp}{88bp}}{rgb(0bp)=(1,1,1);
	rgb(0bp)=(1,1,1);
	rgb(25bp)=(0.67,0.65,0.65);
	rgb(400bp)=(0.67,0.65,0.65)}


\tikzset {_oxe41wk2q/.code = {\pgfsetadditionalshadetransform{ \pgftransformshift{\pgfpoint{89.1 bp } { -108.9 bp }  }  \pgftransformscale{1.32 }  }}}
\pgfdeclareradialshading{_1e90e0moy}{\pgfpoint{-72bp}{88bp}}{rgb(0bp)=(1,1,1);
	rgb(0bp)=(1,1,1);
	rgb(25bp)=(0.67,0.65,0.65);
	rgb(400bp)=(0.67,0.65,0.65)}
\tikzset{every picture/.style={line width=0.75pt}} 

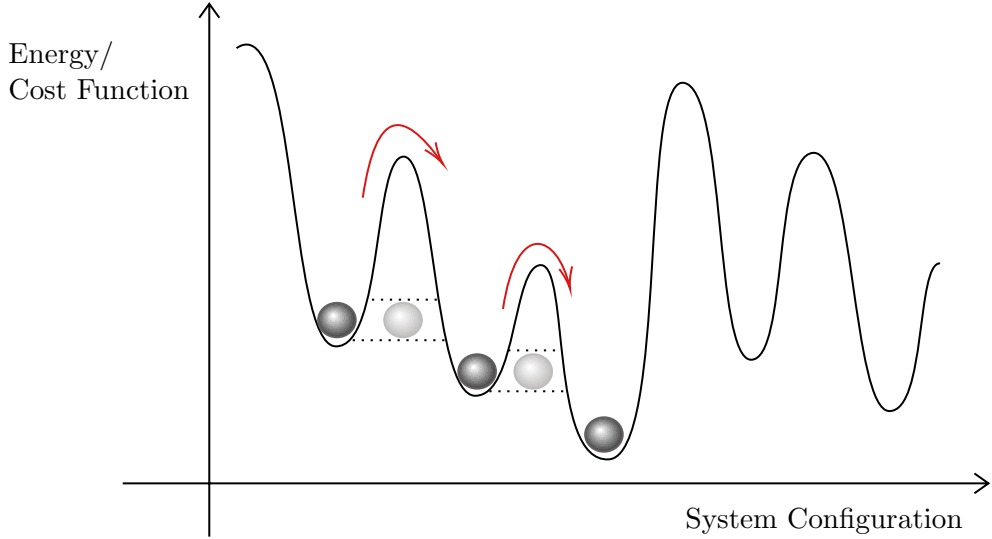
\begin{figure}[h]
	\centering
	\begin{tikzpicture}[x=0.75pt,y=0.75pt,yscale=-1,xscale=1]
	\path (125,325); 
	
	\draw    (235,36.95) .. controls (265.47,15.88) and (260.43,186.1) .. (284.43,186.95) .. controls (308.43,187.81) and (302.97,91.52) .. (318.43,91.52) .. controls (333.89,91.52) and (333,211.81) .. (354.14,211.81) .. controls (375.29,211.81) and (372.69,145.79) .. (387,146.1) .. controls (401.31,146.4) and (391,245.24) .. (420.71,243.81) .. controls (450.43,242.38) and (436.21,52.96) .. (457.86,54.38) .. controls (479.5,55.81) and (473.33,193.69) .. (491.88,193.69) .. controls (510.43,193.69) and (502.7,91.05) .. (522.8,89.62) .. controls (542.9,88.2) and (542.34,219.51) .. (560.89,219.51) .. controls (579.44,219.51) and (574.23,144.57) .. (586.2,145.22) ;
	\path  [shading=_345n2kpid,_etbxlmyj4] (275.73,173.73) .. controls (275.73,168.95) and (279.93,165.08) .. (285.11,165.08) .. controls (290.29,165.08) and (294.49,168.95) .. (294.49,173.73) .. controls (294.49,178.51) and (290.29,182.38) .. (285.11,182.38) .. controls (279.93,182.38) and (275.73,178.51) .. (275.73,173.73) -- cycle ; 
	\draw  [color={rgb, 255:red, 99; green, 95; blue, 95 }  ,draw opacity=1 ][line width=0.75]  (275.73,173.73) .. controls (275.73,168.95) and (279.93,165.08) .. (285.11,165.08) .. controls (290.29,165.08) and (294.49,168.95) .. (294.49,173.73) .. controls (294.49,178.51) and (290.29,182.38) .. (285.11,182.38) .. controls (279.93,182.38) and (275.73,178.51) .. (275.73,173.73) -- cycle ; 
	
	\draw  (178.19,255.87) -- (610.33,255.87)(221.4,14.67) -- (221.4,282.67) (603.33,250.87) -- (610.33,255.87) -- (603.33,260.87) (216.4,21.67) -- (221.4,14.67) -- (226.4,21.67)  ;
	\path  [shading=_4bkrn46oe,_m5kxfk23u] (345.73,199.59) .. controls (345.73,194.81) and (349.93,190.94) .. (355.11,190.94) .. controls (360.29,190.94) and (364.49,194.81) .. (364.49,199.59) .. controls (364.49,204.36) and (360.29,208.24) .. (355.11,208.24) .. controls (349.93,208.24) and (345.73,204.36) .. (345.73,199.59) -- cycle ; 
	\draw  [color={rgb, 255:red, 99; green, 95; blue, 95 }  ,draw opacity=1 ][line width=0.75]  (345.73,199.59) .. controls (345.73,194.81) and (349.93,190.94) .. (355.11,190.94) .. controls (360.29,190.94) and (364.49,194.81) .. (364.49,199.59) .. controls (364.49,204.36) and (360.29,208.24) .. (355.11,208.24) .. controls (349.93,208.24) and (345.73,204.36) .. (345.73,199.59) -- cycle ; 
	
	\path  [shading=_t493ccfgj,_5stw8zhpc] (409.02,231.3) .. controls (409.02,226.53) and (413.22,222.65) .. (418.4,222.65) .. controls (423.58,222.65) and (427.78,226.53) .. (427.78,231.3) .. controls (427.78,236.08) and (423.58,239.95) .. (418.4,239.95) .. controls (413.22,239.95) and (409.02,236.08) .. (409.02,231.3) -- cycle ; 
	\draw  [color={rgb, 255:red, 99; green, 95; blue, 95 }  ,draw opacity=1 ][line width=0.75]  (409.02,231.3) .. controls (409.02,226.53) and (413.22,222.65) .. (418.4,222.65) .. controls (423.58,222.65) and (427.78,226.53) .. (427.78,231.3) .. controls (427.78,236.08) and (423.58,239.95) .. (418.4,239.95) .. controls (413.22,239.95) and (409.02,236.08) .. (409.02,231.3) -- cycle ; 
	
	\path  [shading=_31uveu6mu,_4cx8zkr77] (308.73,173.73) .. controls (308.73,168.95) and (312.93,165.08) .. (318.11,165.08) .. controls (323.29,165.08) and (327.49,168.95) .. (327.49,173.73) .. controls (327.49,178.51) and (323.29,182.38) .. (318.11,182.38) .. controls (312.93,182.38) and (308.73,178.51) .. (308.73,173.73) -- cycle ; 
	\draw  [color={rgb, 255:red, 190; green, 190; blue, 190 }  ,draw opacity=1 ][line width=0.75]  (308.73,173.73) .. controls (308.73,168.95) and (312.93,165.08) .. (318.11,165.08) .. controls (323.29,165.08) and (327.49,168.95) .. (327.49,173.73) .. controls (327.49,178.51) and (323.29,182.38) .. (318.11,182.38) .. controls (312.93,182.38) and (308.73,178.51) .. (308.73,173.73) -- cycle ; 
	
	\path  [shading=_1e90e0moy,_oxe41wk2q] (373.73,199.59) .. controls (373.73,194.81) and (377.93,190.94) .. (383.11,190.94) .. controls (388.29,190.94) and (392.49,194.81) .. (392.49,199.59) .. controls (392.49,204.36) and (388.29,208.24) .. (383.11,208.24) .. controls (377.93,208.24) and (373.73,204.36) .. (373.73,199.59) -- cycle ; 
	\draw  [color={rgb, 255:red, 190; green, 190; blue, 190 }  ,draw opacity=1 ][line width=0.75]  (373.73,199.59) .. controls (373.73,194.81) and (377.93,190.94) .. (383.11,190.94) .. controls (388.29,190.94) and (392.49,194.81) .. (392.49,199.59) .. controls (392.49,204.36) and (388.29,208.24) .. (383.11,208.24) .. controls (377.93,208.24) and (373.73,204.36) .. (373.73,199.59) -- cycle ; 
	
	\draw  [dash pattern={on 0.84pt off 2.51pt}]  (302.6,163.47) -- (337,163.47) ;
	\draw  [dash pattern={on 0.84pt off 2.51pt}]  (293,183.87) -- (339,183.87) ;
	\draw  [dash pattern={on 0.84pt off 2.51pt}]  (370.56,188.98) -- (396.56,188.98) ;
	\draw  [dash pattern={on 0.84pt off 2.51pt}]  (361.22,209.56) -- (399.76,209.56) ;
	
	\draw [color={rgb, 255:red, 220; green, 21; blue, 21 }  ,draw opacity=1 ]   (298,112.17) .. controls (304.86,66.11) and (318.44,68.07) .. (336.87,93.58) ;
	\draw [shift={(338,95.17)}, rotate = 234.87] [color={rgb, 255:red, 220; green, 21; blue, 21 }  ,draw opacity=1 ][line width=0.75]    (10.93,-3.29) .. controls (6.95,-1.4) and (3.31,-0.3) .. (0,0) .. controls (3.31,0.3) and (6.95,1.4) .. (10.93,3.29)   ;
	\draw [color={rgb, 255:red, 220; green, 21; blue, 21 }  ,draw opacity=1 ]   (368,168.17) .. controls (374.86,122.11) and (395.17,130.79) .. (401.14,156.5) ;
	\draw [shift={(401.49,158.1)}, rotate = 258.48] [color={rgb, 255:red, 220; green, 21; blue, 21 }  ,draw opacity=1 ][line width=0.75]    (10.93,-3.29) .. controls (6.95,-1.4) and (3.31,-0.3) .. (0,0) .. controls (3.31,0.3) and (6.95,1.4) .. (10.93,3.29)   ;
	
	\draw (119.71,32.09) node [anchor=north west][inner sep=0.75pt]   [align=left] {Energy/\\Cost Function};
	\draw (457.67,266.75) node [anchor=north west][inner sep=0.75pt]   [align=left] {System Configuration};
	
	\end{tikzpicture}
	
	\caption{Arrows describe how SA climbs over the valleys through thermal fluctuations while QA uses tunneling effect to move through them as depicted by circles.}
	\label{fig:qa}
\end{figure}

Quantum annealing is experimentally realizable on commercially available D-Wave quantum processing units (QPUs) \cite{johnson2011quantum} and sparked the interest to solve combinatorial optimization problems,  with applications ranging from transportation problems \cite{neukart2017traffic,domino2021quadratic}, finance \cite{rebentrost2018quantum}, chemistry \cite{perdomo2012finding}, to scheduling problems \cite{venturelli2015quantum}.

\subsubsection{Encoding the Problem: The Ising Model and QUBO}
In order to describe how quantum annealers can be used for solving problems, it is necessary to know how to formulate and encode them in such a way that can be processed by quantum machines. For this, one needs to know how to map the elements in our search space to certain states, and to energies.

In quantum mechanics, the energy configuration of the system is driven by the Hamiltonian: an operator that describes the forces to be applied to a single qubit and qubit pairs to move the state of the system into some desired state. The system is initialized with the ground state of a Hamiltonian $H_0$ whose ground state is easy to prepare. The problem Hamiltonian $H_P$ is introduced gradually to the system, whose ground state encodes the solution to the problem of interest. And then, if $H_{0}$ and $H_{P}$ do not commute and the evolution is slow enough, according to the Quantum Adiabatic Theorem \cite{farhi2000quantum}, during an annealing time $T$, the system will remain, ideally, in the minimum energy state throughout the process. In other words, the evolution of the system is described by
\begin{equation}
H(t)=\left(1-\frac{t}{T}\right) H_{0}+\frac{t}{T} H_{\mathrm{P}},
\end{equation}
so that initialized with the ground state of $H_0$ at time $t=0$, we obtain the solution to our problem measuring the quantum state at a time $t=T$. Therefore, to solve the problem of interest, we need to design $H_P$, the problem Hamiltonian. Then, we use the properties of nature as described by Quantum Adiabatic Theorem, to solve the problem.

Consider particles arranged on the nodes of a graph, where each particle can interact with its neighbors. The particles can be either in state $-1$ or $+1$, called the spin. An assignment of $-1$s and 1s to the spins is known as the spin configuration. A mathematical model called the Ising Model is used for describing the properties of such physical systems that evolve in time. The interaction force or the coupling strength between the particles is denoted by $ J_{ij} $, and an external force $h_i$ called the qubit bias is applied on each particle. The energy of a configuration is given by
\begin{equation}
    H(s)=\sum_{i} h_{i} s_{i}+\sum_{i<j} J_{i j} s_{i} s_{j},
\end{equation}
where $s_i \in \{-1,+1\}$. 
 
D-Wave quantum annealers also fit this setting. The problem Hamiltonian $H_P$ is encoded through an Ising Model,  simply by substituting $s_i$ by the Pauli-$z$ operator $\sigma_{i}^{z}$ acting on $i$th qubit:
\begin{equation}
    H_{F}=\sum_{i} h_{i} \sigma_{i}^{z}+\sum_{i<j} J_{i j} \sigma_{i}^{z} \sigma_{j}^{z}.
\end{equation}
The Hamiltonian $H_0$ for initializing the system is the transverse field Hamiltonian
\begin{equation}
    H_{0}=-h_{0} \sum_{i=1}^{N} \sigma_{i}^{x},
\end{equation}
where $\sigma_{i}^{x}$ is the Pauli-$x$ operator acting on the $i$th qubit
so that the ground state of $H_{0}$ is the equal superposition of all basis states. One should note that the problem of finding the spin configuration which minimizes $H(s)$ is NP-Hard in general \cite{Barahona_1982}.  For further information regarding quantum annealing, we refer the readers to \cite{mcgeoch2014adiabatic}.

The significance of the ability of the QUBO model to encompass many problems in combinatorial optimization is enhanced by the fact that the QUBO model can be shown to be equivalent to the Ising model. The transformation between QUBO and Ising model can be performed easily using the mapping $x_{i} \leftrightarrow \frac{1-s_{i}}{2}$. Hence any problem for which QUBO formulation is known, or can be expressed using QUBO formulation, can be attempted in principle to be solved using quantum annealing. We would like to remark that it is often more desirable to work with QUBO formulation than Ising Model. 
                     
\subsubsection{Running Problems on D-Wave}

The D-Wave quantum processing unit (QPU) is a lattice of interconnected qubits. While some qubits connect to others via couplers, the D-Wave QPU is not fully connected. Therefore, the variables can not be mapped directly to the physical qubits on the machine. Hence, each variable is represented by a set of qubits called the chain, and the qubits in a chain are coupled strongly enough based on a parameter called chain strength so that they end up in the same state. The existence of longer chains increases the error in the results. This process of mapping the variables to the physical qubits is known as the minor-embedding problem and can be handled by D-Wave. Nevertheless, an increased chain length reduces the quality of the solutions obtained.

When submitting problems to D-Wave, certain parameters need to be tuned besides the chain strength. This includes the number of reads and the annealing time. The number of reads is the total number of samples returned by D-Wave. Ideally, the sample with the lowest energy corresponds to the spin/qubit configuration of the ground state. However, due to the limitations of the device and the external noise, the ground state is not always achieved. 

Annealing time depends both on the problem and the problem instance that is in consideration. There is a huge effort to detect whether quantum annealing provides any speedup against classical methods \cite{ronnow2014defining, hen2015probing, katzgraber2015seeking, mandra2018deceptive}, and there are both supporting and opposing claims in this regard. Anyhow, it is still a field that is expanding and promising with the emergence of quantum-classical hybrid methods \cite{141055aa82:online}.

%% file: markov.tex
\subsection{Markov Random Fields}
Now, we will explore another tool that can be used to model our problems when using quantum annealing.

\subsubsection{Markov Chains}\label{sec: markovchains}

Suppose that we want to predict the weather of the following day. Is it more likely that tomorrow will be cold? Or perhaps, will it be rainy? To solve the problem, we could make the following analysis: let us say that the probability of a cold day in our city is 1/4; if today it is cold, the likelihood of tomorrow being cold would be the joint probability of these events, that is, $P(\text{today is cold and tomorrow is cold}) = \frac{1}{4} \cdot \frac{1}{4} = \frac{1}{16}
$. This probability seems too low, and usually, cold days follow one another in a row. If today is the first cold day in a while, a probability of 1/16 for tomorrow being also a cold day does not reflect our experience entirely.

Markov's revolutionary idea was to approach this problem in a different way; instead of viewing each probability as independent from one day to another, he decided to consider the present state as a necessary piece of information, the only one actually, for calculating the weather of the next day. In our analogy, this is like saying that the probability of being cold tomorrow depends only on the state of the weather of today. If today it is cold, the probability of tomorrow being cold should be higher than it would be if today is too hot instead. With this approach, a sequence of weather states can give rise to a chain of events that satisfies what it is called the \textit{Markovian property}.

Consider a discrete-time stochastic process, this is, a sequence of random variables $\{X_{n\geq0}\}$, with state space $E$. Here, we will consider a countable state space. If for all integers $n\geq 0$ all the states $i_0, i_1, ..., i_{n-1}, i, j$ satisfy
\begin{equation} \label{eq:mc}
P(X_{n+1}=j\vert X_n=i, X_{n-1}= i_{n-1}, \dots, X_0=i_0)= P(X_{n+1}=j\vert X_n=i)
\end{equation}
we say that the process is a Markov Chain. 

In our example, the state space has three states, namely $\{\text{hot, cold, rainy}\}$. As for the conditional probabilities $ P_{i,j}\equiv P(X_{n+1}=j\vert X_n=i)$ (the probability of $j$ happening tomorrow given that $i$ occurs today), it is customary to arrange them in a matrix called the \textit{transition matrix} as depicted in Table \ref{table: markov}.  For instance, $P_{\text{hot,cold}} = 1/4$ represents the probability of a cold day happening tomorrow, given that a hot day occurs today.

\begin{table}[h!]
	\centering
	\begin{tabular}{ |c|ccc| } 
		\hline
		& hot & cold & rainy \\
		\hline
		hot  & 1/2 & 1/4 & 1/4 \\ 
		cold & 1/6 & 1/2 & 1/3 \\ 
		rainy  & 1/8 & 3/8 & 1/2 \\ 
		\hline
	\end{tabular}
	\caption{Transition probabilities for the Markov Chain with the state space $\{$hot, cold, rainy$\}$.}
	\label{table: markov}
\end{table}

\begin{figure}[h!]
\begin{center}
\includegraphics[scale=0.25]{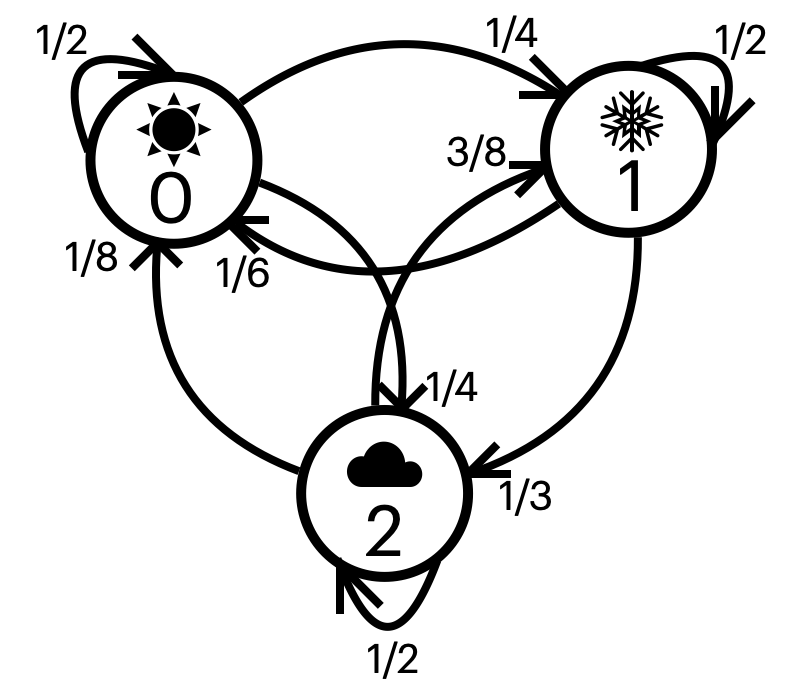}
\caption{The graph representation for the Markov Chain with state space hot day, cold day, rainy day. Each transition is labeled with the appropriate probability.}
\label{fig:markov}
\end{center}
\end{figure}

A transition matrix $P$ can also be visualized as a graph $G$ whose vertices represent the states of $E$. This graph has an oriented edge from $i$ to $j$ that will appear labeled with the probability $P_{i,j}$. The graph representation for our example is given in Fig. \ref{fig:markov} 

\subsubsection{Markov Random Fields}
Let us generalize the concept of a Markov Chain and introduce the Markov Random Field. A \textit{Markov Random Field} (MRF) or \textit{Markov Network} is a set of random variables that form an undirected graph which obeys the Markovian property \cite{spitzer1971markov, preston1973generalized}. Random variables are represented by the nodes of the network (the vertices of the graph). Each random variable can take values from a finite set, and an assignment to the variables is called a \textit{configuration}. Any pair of random variables that are not connected are conditionally independent. 

\begin{figure}[h!]
\begin{center}
\includegraphics[scale=0.27]{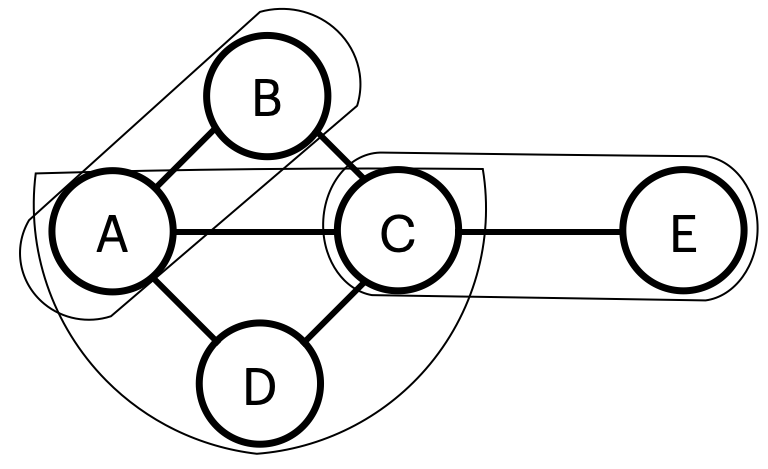}
\caption{An example of a Markov Random Field.}
\label{fig:mrf}
\end{center}
\end{figure}

In the example visualized in Fig. \ref{fig:mrf}, we can observe several fully connected subgraphs of the graph. We denote these subgraphs as \textit{cliques}. A \textit{potential} function is defined over the cliques of the graph, assigning a value to every configuration of the clique. It is important to note though that, these values need not represent actual probabilities, that is, they may not add up to 1.

Concerning the potential functions, these are associated with an energy function. The total energy is defined as the sum over all clique potentials. High energy will indicate a low probability for a certain configuration to happen, and conversely, low energy will suggest that a specific configuration has more chances to occur. Although we do not go into the details of the energy calculation here, we refer interested readers to \cite{Kindermann1980MarkovRF}.

From Fig.\ref{fig:mrf} let us choose the clique $C_{0}=\{C,E\}$ with $C\in\{0,1\}$ and $E\in\{0,1\}$, that is $C$ and $E$ are binary random variables. We can define the potentials for the different configurations of the clique as shown in Table \ref{table:potentials}.

\begin{table}[h!]
\centering
\begin{tabular}{ |c|c|c| } 
\hline
$C$ & $E$ & $p(C,E)$ \\
\hline
0 & 0 & 0.3 \\ 
\hline
0 & 1 & 0.9 \\
\hline
1 & 0 & 2.6 \\
\hline
1 & 1 & 5.0 \\
\hline
\end{tabular}
\caption{Potentials for the clique $C_{0}=\{C,E\}$ from Fig. \ref{fig:mrf}. }
\label{table:potentials}
\end{table}

We will be primarily interested in the configuration with the highest probability of occurrence, equivalently the lowest energy. This is where the idea of \textit{optimization} comes into play. The idea here is to encode the states of interest as the nodes of the network and then define the potentials so that using an optimization algorithm the lowest energy configuration is obtained.

\subsubsection{Quantum Annealing and Markov Random Fields}

Markov Random Fields are inspired by the Ising Model \cite{Kindermann1980MarkovRF} and can be viewed as an attempt for generalization of the 2-local Ising Model traditionally defined over a lattice. From the view of probability theory, the probability that each particle ends up in state -1 or 1 depends only on the spin values of its neighbors, hence satisfying the Markovian property. Therefore, the Ising model is indeed a special case of MRF, where there are cliques of sizes 1 and 2 only, and the state space is $\{-1,1\}$.

As mentioned previously in this chapter, quantum annealing provides a way to take advantage of the properties of a quantum system to find solutions for optimization problems. Through the use of tools such as the Ocean software development kit, one can create a Markov Network and use D-Wave to sample configurations of the random variables with the lowest energy. This is accomplished by D-Wave, by creating a QUBO model from the potentials of the Markov Network as described below. 

Given random variables $X_1$ and $X_2$ with potentials $\phi_{00},\phi_{01},\phi_{10},\phi_{00}$ where $\phi_{ij}$ is the potential for $X_1=i$ and $X_2=j$, the corresponding QUBO formulation is defined as 
\begin{equation}
    (\phi_{10} - \phi_{00}) x_1 + (\phi_{10} - \phi_{00}) x_2 + (\phi_{11} - \phi_{10} - \phi_{01} + \phi_{00}) x_1x_2 + \phi_{00}.
\end{equation}

%% file: music_optimization.tex
Computational music generation is a field that emerged with the invention of computers in the late 1940s. The first computer to play a music piece was CSIR Mark-1 \cite{doornbusch2017early}. In the initial period, most studies focused on playing an existing music piece using computers rather than composing a new music piece. Later with the increase of computational power, the focus of researchers shifted to creating new music with the help of 
computers \cite{bogdanov2001all}. Programs and set of programming languages known as MUSIC-N were developed by Max Mathews at Bell Laboratories in 1957 \cite{manning1993computer}. Thus the field of computational music, which is also known as algorithmic music, began \cite{ShortHis65:online}. Various tools and techniques for algorithmic music generation has been extensively discussed in \cite{miranda2001composing}.

Optimization is a widely used technique for computational music generation. As mentioned previously, any piece of music can be seen as a sequence of sounds and rests. These sequences can be identified for their adherence to a particular music style or any other musical property in such a way that if a musical element deviates from that musical style or property, then having that musical element in the generated music piece is associated with a cost or penalty. Therefore through the optimization process,  a sequence of music elements with minimum cost, i.e., one minimizing the deviation from the pre-identified music style or any other musical property, can be generated.

Now, we will briefly summarize some of the techniques used for computational music generation, where the music generation is achieved through optimizing some parameters related to the music piece. 

Statistical modeling is a common technique used in computational music generation. Under this technique, the existing music corpus is analyzed and its statistical properties are derived. The new piece of music starts with a given note. Then, the probabilities of having each possible note (pitch) as the next note in generated music piece are calculated. The note with the highest probability is selected as the next note in the new music piece. Thus, sequentially the entire new music piece is obtained. Here it is important to note that the newly generated music piece follows the statistical properties of the music piece from which the probabilities were calculated \cite{manaris2007corpus}. \textit{Illiac Suite} is considered to be the first musical score generated by a computer in 1957 \cite{hiller1979experimental,funk2016zen}, using a statistical method known as Markov Chains. Markov Chains as a tool for music generation has been described in detail in Sec \ref{sec: markovchains}. It is an active area of research even to this day \cite{conklin2003music}.

Constraint programming is a programming paradigm where instead of steps to solve a problem, the properties of the solution are specified. These properties are the constraints that the solution is expected to follow. Mainly, it is an expansion of Constraint Logic Programming, which in turn is an expansion of Satisfiability Problem (SAT) \cite{apt2003principles}. The set of constraints with a general-purpose search algorithm can solve large practical combinatorial and scheduling problems \cite{van1999cplan}. Since music composition is a process where at every step, the choices available for the composer are in the form of combinations of notes, chords,  or intervals; these combinations are constrained naturally by the rules of music, such as melody or harmony generation rules \cite{truchet2001constraint}. This resemblance in the process of music generation and constraint programming has been used in modeling and generation of various musical forms such as counterpoint, harmony, rhythm, form, and instrumentation   \cite{anders2018compositions}.  

Genetic Algorithm (GA) has been claimed to simulate the creative process of music composition through its operators such as \textit{mutation, selection} and \textit{crossover}. GA is an optimization technique inspired by Darwin's theory of natural selection and his observation that in any ecosystem, the species' growth is governed through a process where only the fittest offsprings survive among a huge population, and the surviving ones reproduce and form the next generation. The process of reproduction and the natural selection continues and gives rise to different species. Accordingly, the first step to use GA for music composition is to generate a set (a.k.a \textit{population} in GA terminology) of random solutions. Then the set is updated using the technique of offspring creation or \textit{reproduction} such as \textit{mutation, selection} and \textit{crossover}, and only the potentially feasible solutions which are highly favored based on certain \textit{fitness function} are kept. The cycle of \textit{reproduction} and keeping only the fittest members of the population continues and a \textit{solution} is generated
 \cite{biles2007evolutionary}. 

GA was used in computer-based music generation first in the early 1990's \cite{horner1991genetic}. NEUROGEN was a GA-based program developed in 1991 to produce and combine several musical fragments \cite{140338}. Lee Spector and Adam Alpern also used GA to identify deep musical structures and generate music \cite{spector1995induction}. MusicGenie, developed in 2006, was a music composition program inspired by GA  \cite{samadani2007music}. 


Machine Learning is the latest addition in the quiver of optimization-based techniques for music generation. Neural Network based machine learning methods have been used for music composition since 1988 \cite{todd1988sequential}. As the subsequent notes/chords or any music element depends on the previous music elements in the music sequence, the Recurrent Neural Network and Long Short Term Memory Networks had been among the favourite choices of neural network researchers for music composition \cite{MARINESCU2019117, kotecha2018generating}.

In the cutting edge of computational music generation techniques, Deep Learning is among the most successful and recent ones. Briot, Jean-Pierre, and Hadjeres have provided a detailed survey of deep learning based methods for music generation \cite{briot2020deep}.  Using MuseNet, a deep neural network, a 4-minute long music piece with ten different instruments is composed. MuseNet also combines several different musical styles, including those of Mozart, Beatles, and Country music \cite{MuseNet3:online}. These techniques are being further refined with the use of Generative Adversarial Networks (GAN) and the development of General Purpose Transformers (GPT-2 and GPT-3) \cite{park2021review}.

Integer Linear Programming (ILP), which is often referred to as Integer Programming (IP) in the literature, is another main framework for solving combinatorial problems. Though it shares a lot in common with Constraint Programming, IP only uses variables with integer values to represent the problem as described in the previous subsection. Based on our knowledge, integer programming has not been used extensively in the literature in the scope of music generation. Some works in this line are the following. The natural one-to-one mapping between integers and pitch values of the 12-tonal music system has been exploited by Tsubasa Tanaka et al. \cite{tanaka2016integer} for musical motif analysis of existing masterpieces of music. Nailson dos Santos Cunha et al. have generated guitar solo music using integer programming \cite{cunha2018generating}.

The bottleneck in the ILP/IP approach of solving a combinatorial problem is the method of optimization used.  Therefore several heuristic methods such as Hill Climbing, Simulated Annealing, Reactive Search Optimization, Ant Colony Optimization, Hopfield Neural Networks, Tabu Search have been used to reach the solution set \cite{conforti2014integer, glover1990tabu}. The next imperative step would be using quantum methods for solving IP problems. With this motivation, Integer Programming becomes especially important in the context of this chapter as any ILP formulation can be converted into QUBO formulation as described in Sec. \ref{sec: ilptoqubo}.

%% file: music_generation.tex
In this section, we will explore how one can use quantum annealing to compose different elements of music such as melody, rhythm, and harmony. We will describe how one can model the process of music creation through a set of constraints and an objective function. The codes used in this section to generate the music pieces are available in the form of a Jupyter notebook on \url{https://doi.org/10.5281/zenodo.5856930}.

\subsection{Melody Generation} \label{sec: melody}

As described in the previous section, quantum annealing aims to find the optimal solution to an optimization problem. We will start by investigating different ways of formulating the process of melody generation as an optimization problem. When we talk about melody generation, we refer to the generation of the pitches only. 

\subsubsection{Model} \label{sec: model}
Suppose that we want to generate a melody consisting of $n$ notes, where the pitches belong to the set $P=\{p_1,p_2,\dots,p_k \}$. We will define the binary variables $x_{i,j}$ for $i \in [n]$ and $j \in P$ as
\begin{equation}\label{eq:model}
x_{i,j} =  \begin{cases}%
1      & \text{note at position $i$ is $j$,}\\
0 & \text{otherwise.}
\end{cases}
\end{equation}
The total number of variables required in this formulation is $n|P|$. Note that we aim to generate $n$ notes simultaneously. At the end of the optimization process, we will obtain an assignment to the binary variables $x_{i,j}$, which will indicate the pitch of the note selected for each position. 

Next, we will define some constraints which can be then included in the objective function using the penalty method described in Section \ref{sec:optimization}. The first rule we need to incorporate is that only one of the variables $x_{i,j}$ is equal to 1 for each position $i$. The rule is necessary as exactly one pitch should be selected for each time point. This is equivalent to having the following constraint for each $i \in [n]$:
\begin{equation}\label{eq:single_note}
    \sum_{j \in P} x_{i,j} = 1.
\end{equation}
This constraint is the backbone of our formulation, and it should be included in the QUBO using a sufficiently large penalty coefficient as it should never be violated. 

Let us go through an example. Let $P=\{\mathtt{C4}, \mathtt{D4}, \mathtt{E4}, \mathtt{G4} \}$ and $n=5$. The QUBO formulation has 20 variables defined through Eq.~\ref{eq:model} and 5 penalty terms of the form
\begin{equation*}
    \left ( 1- \sum_{j \in P} x_{i,j} \right )^2
\end{equation*}
for each $i$, obtained from Eq~\ref{eq:single_note}. At this point, any sequence of $5$ notes have 0 energy and is equally likely to be produced and there are in total $4^5$ such sequences. The list of first 5 samples obtained from D-Wave as a result of running the QUBO formulation described above is given in Table \ref{table:samples}. 

\begin{table}[]\centering
\begin{tabular}{|l|cccc|cccc|cccc|cccc|cccc|}
\hline
                  & \multicolumn{4}{l|}{$i=1$} & \multicolumn{4}{l|}{$i=2$} & \multicolumn{4}{l|}{$i=3$} & \multicolumn{4}{l|}{$i=4$} & \multicolumn{4}{l|}{$i=5$} \\ \hline
\textbf{Sample 1} & 0 & 0 & 1 & 0 & 0 & 0 & 1 & 0 & 0 & 0 & 0 & 1 & 0 & 1 & 0 & 0 & 0 & 1 & 0 & 0 \\ \hline
\textbf{Sample 2} & 0 & 1 & 0 & 0 & 0 & 0 & 0 & 1 & 1 & 0 & 0 & 0 & 1 & 0 & 0 & 0 & 0 & 1 & 0 & 0 \\ \hline
\textbf{Sample 3} & 0 & 1 & 0 & 0 & 0 & 1 & 0 & 0 & 0 & 0 & 0 & 1 & 0 & 0 & 0 & 1 & 0 & 0 & 1 & 0 \\ \hline
\textbf{Sample 4} & 0 & 0 & 1 & 0 & 0 & 0 & 0 & 1 & 0 & 0 & 1 & 0 & 1 & 0 & 0 & 0 & 0 & 0 & 0 & 1 \\ \hline
\textbf{Sample 5} & 0 & 1 & 0 & 0 & 0 & 0 & 1 & 0 & 1 & 0 & 0 & 0 & 0 & 0 & 1 & 0 & 0 & 1 & 0 & 0 \\ \hline
\end{tabular}
\caption{The list of first 5 samples obtained from D-Wave. The columns represent the values of the variables $x_{1,C4},x_{1,D4}, x_{1,E4}, x_{1,G4}, x_{2,C4}, \dots, x_{5,G4} $ in the given order. }
\label{table:samples}
\end{table}

The energies of all the samples in Table \ref{table:samples} are 0 as no constraint is violated, i.e., precisely one of the variables is 1 for each $i=0$. The resulting note sequences are given in Table \ref{table:notes} 

\begin{table}[h]
	\centering
	\begin{tabular}{|l|l|}
		\hline
		\textbf{Sample 1} & $\mathtt{E4}-\mathtt{E4}-\mathtt{G4}-\mathtt{D4}-\mathtt{D4}$ \\ \hline
		\textbf{Sample 2} & $\mathtt{D4}-\mathtt{G4}-\mathtt{C4}-\mathtt{C4}-\mathtt{D4}$ \\ \hline
		\textbf{Sample 3} & $\mathtt{D4}-\mathtt{D4}-\mathtt{G4}-\mathtt{G4}-\mathtt{E4}$ \\ \hline
		\textbf{Sample 4} & $\mathtt{E4}-\mathtt{G4}-\mathtt{E4}-\mathtt{C4}-\mathtt{G4}$ \\ \hline
		\textbf{Sample 5} & $\mathtt{D4}-\mathtt{E4}-\mathtt{C4}-\mathtt{E4}-\mathtt{D4}$ \\ \hline
	\end{tabular}
	\caption{The list of note sequences corresponding to the samples given in Table \ref{table:samples}.}
	\label{table:notes}
\end{table}

So far, we have only defined a single rule ensuring a single note at each time point. In general, one would like to introduce some more rules while composing a melody as we will discuss next. 

\subsubsection{Rules About Consecutive Notes}

Suppose that we want to add a restriction that the note $p_l$ does not appear after the note $p_k$. This is useful for avoiding particular intervals and amending our model. Such a restriction can be incorporated into the QUBO formulation by adding the term 
\begin{equation}\label{eq:consec} 
x_{i,p_k}x_{i+1,p_l}     
\end{equation}
to the objective function with a suitable penalty $C$ for each $i \in [n-1]$. Note that when both variables equal 1 simultaneously, a penalty of $C$ is added to the objective function.

Alternatively, we can express the same rule using the following constraint:
\begin{equation}\label{eq:constraint}
x_{i,p_k} + x_{i+1,p_l} \leq 1.     
\end{equation}
In this case, the inequality should be first transformed into equality by using slack variables and then added to the objective function.

Now let us consider a rule saying that the same note does not appear more than twice in a row. Similar to what we had above, the term $x_{i,j}x_{i+1,j}x_{i+2,j} $ can be added to the objective. However, this is not a quadratic term, and quadratization is needed to obtain a QUBO. 

Alternatively, the rule can be expressed by the following constraint
\begin{equation}\label{eq:samenote}
x_{i,p_j} + x_{i+1,p_j} + x_{i+2,p_j} \leq 2
\end{equation}
for each $i = 0, \dots ,n-2$, which forces that at most two of the variables are equal to 1 simultaneously. 

Taking the previous example and assuming that $P=\{\mathtt{C4}, \mathtt{D4}, \mathtt{E4}, \mathtt{G4} \}$ and $n=5$, let us also add the rule that $\mathtt{G4}$ does not follow $\mathtt{D4}$ using Eq.~\ref{eq:consec} and include Eq. \ref{eq:samenote} in our formulation as well. The first 5 samples from the experiment results are listed in Table \ref{table:notes2}. All the sequences in the table have 0 energy and obey the rules we have incorporated. 

\begin{table}[h]
\centering
\begin{tabular}{|l|l|}
\hline
\textbf{Sample 1} & $\mathtt{G4}-\mathtt{D4}-\mathtt{E4}-\mathtt{G4}-\mathtt{C4}$ \\ \hline
\textbf{Sample 2} & $\mathtt{E4}-\mathtt{E4}-\mathtt{C4}-\mathtt{E4}-\mathtt{G4}$ \\ \hline
\textbf{Sample 3} & $\mathtt{E4}-\mathtt{D4}-\mathtt{C4}-\mathtt{E4}-\mathtt{G4}$ \\ \hline
\textbf{Sample 4} & $\mathtt{D4}-\mathtt{E4}-\mathtt{G4}-\mathtt{D4}-\mathtt{C4}$ \\ \hline
\textbf{Sample 5} & $\mathtt{D4}-\mathtt{E4}-\mathtt{G4}-\mathtt{D4}-\mathtt{E4}$ \\ \hline
\end{tabular}
\caption{The list of note sequences obtained from D-Wave after incorporating constraints given in Eq.~\ref{eq:consec} and Eq.~\ref{eq:samenote}.
\label{table:notes2}}
\end{table}

\subsubsection{Semitones and Augmented Intervals}
Now let us investigate the different ways we can choose set $P$. We can identify the notes through the number of semitones between the lowest pitch and the pitch in consideration. The binary variables $x_{i,j}$s are defined as
\begin{equation}\label{eq:modelchromatic}
x_{i,j} =  \begin{cases}%
1      & \text{note at position $i$ is $j$ semitones apart from the lowest pitch of the sequence,}\\
0 & \text{otherwise.}
\end{cases}
\end{equation}
for each $i\in[n]$ and $j\in P$. 

When $P$ is selected as $\{0,1,2,\dots,12\}$, then it represents the notes from a chromatic scale. Note that this representation is independent of the key chosen as the result may be interpreted in any key. For instance, if we let $x_{i,0} = \mathtt{C4}$, subsequently the resulting $P$ is the set of notes from the chromatic scale of $\mathtt{C}$.

Identifying the notes through semitones, it will be easier for us to model some rules like avoiding particular intervals. Let $A$ be the list of intervals in semitones, that we would like to avoid. This rule can be incorporated either by adding the following term to the objective function
\begin{equation}\label{eq:augmented}
   x_{i,j}x_{i+1,j'}  
\end{equation}
for all $i \in [n], |j'-j| \in A$ or translates to the following constraint:
\begin{equation}
x_{i,j} + x_{i+1,j'} \leq 1 \text{ for all } i \in [n], |j'-j| \in A.
\end{equation}
so that whenever an unallowed interval is used, we have a penalty. 

What we can do further is to take $P$ as a subset of the chromatic scale. For instance, the set $P=\{0,2,4,5,7,9,11,12\}$ represents the notes from a major scale. 

Let us also assume that we would like to set the first and the last pitch of the generated music piece as the first degree of the scale. This can be incorporated by simply adding the terms
\begin{equation}\label{eq:firstlast}
    (1-x_{1,0}),~~(1-x_{n,0}),
\end{equation}
to the objective.

\begin{figure}[h]
\begin{center}
\includegraphics[scale=0.6]{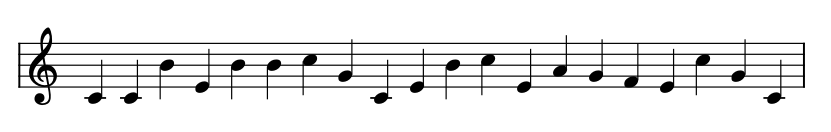}
\caption{The sequence of semitones is obtained from D-Wave hybrid solver and it is interpreted in $\mathtt{C~Major}$. }
\label{fig:augmented}
\end{center}
\end{figure}

Let us go over an example. We will let $A=[6,8,10,12]$, so that we would like to avoid the intervals tritone, augmented fifth, augmented sixth, and augmented seventh. Assuming that $P=\{0,2,4,5,7,9,11,12\}$, $n=20$ and incorporating rules Eq.~\ref{eq:single_note}, Eq.~\ref{eq:samenote}, Eq.~\ref{eq:augmented}, and Eq~\ref{eq:firstlast}, we obtain a QUBO formulation. Due to the increased number of constraints, the solution returned by D-Wave QPU violates some constraints and fails to return the ground state. Using D-Wave hybrid solver, the obtained melody interpreted in $\mathtt{C~Major}$ is given in Figure \ref{fig:augmented}. Note that not all the constraints are satisfied in this sample as well. This can be viewed both as a caveat and a feature, as the violation of some constraints introduces some randomness to the process.

s

\subsubsection{Diatonic Scale and Tendency Notes }
Previously, we investigated how one can identify the notes through semitones. All music pieces have a definite key signature. Therefore it is often more suitable to work with pitches from a particular scale.

For simplicity, let us consider the 8 degrees of a diatonic scale. In this case, the set $P$ consists of $d_1,d_2,\dots,d_8$ where $d_j$ is the $j$'th degree of the scale. Instead of defining the variables by the pitches, we will define them through degrees as in Eq. \ref{eq:modeldiatonic}:
\begin{equation}\label{eq:modeldiatonic}
x_{i,j} =  \begin{cases}%
1      & \text{note at position $i$ is $d_j$,}\\
0 & \text{otherwise.}
\end{cases}
\end{equation}
for each $i\in[n]$ and $d_j\in P$. 

As a result of the optimization procedure, we will obtain a degree sequence, which can then be translated into a note sequence based on the chosen scale. Hence, our model is readily adaptable to different scales. Furthermore, the rules described in the previous subsections are still applicable. 

Some notes in the scale are less stable than the others which are known as the \textit{tendency notes} and tend to resolve to the stable ones. Let us examine how to incorporate rules about tendency notes.  According to the rule, degrees 2,4, and 6 resolve down by one step and degree 7 resolves to the octave. To reflect the rule about tendency notes, for each $i \in [n-1]$, we will add the terms
\begin{equation}\label{eq:tendency}
x_{i,2}(1-x_{i+1,1}),~~
x_{i,4}(1-x_{i+1,3}),~~
x_{i,6}(1-x_{i+1,5}),~~
x_{i,7}(1-x_{i+1,8}),
\end{equation}
to the objective.

\begin{figure}[h]
\begin{center}
\includegraphics[scale=0.6]{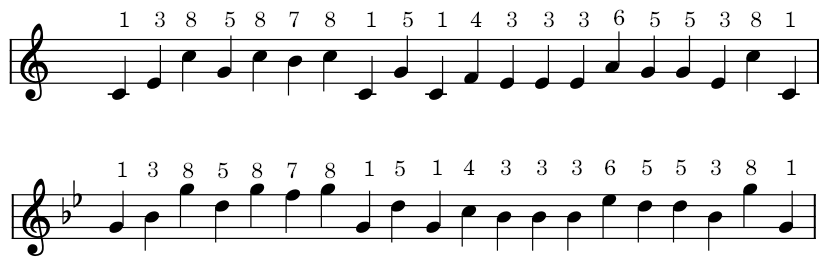}
\caption{The degree sequence is obtained from D-Wave and it is interpreted in $\mathtt{C~Major}$  and $\mathtt{G~Minor}$ (natural) in the presented music scores, respectively. }
\label{fig:degree}
\end{center}
\end{figure}

Using \cref{eq:firstlast,eq:modeldiatonic,eq:tendency} to formulate our model and setting $n=20$, the degree sequence obtained from D-Wave corresponding to one of the samples with the lowest energy is given in Figure \ref{fig:degree}, together with interpretations for different scales.

\subsubsection{Objective Function}

Having discussed how to incorporate different constraints into the model, we can now explore how we can modify the objective function to differentiate between the feasible solutions. When we have only the constraints, all sequences of notes that don't violate any of the constraints are feasible solutions, and they are equally likely to be sampled since they have the lowest possible energy. Any violated constraint contributes to the energy value by the penalty value of the constraint. Note that as mentioned earlier, sometimes we would like particular constraints to be never violated (like Eq.~\ref{eq:single_note}), while a solution in which some constraints are violated can still be desirable.

In order to differentiate between the feasible solutions, we can give some ``rewards'' to particular sequence of notes, i.e. decrease their energy. For instance,  we might give a higher reward for pitch $\mathtt{D4}$ following $\mathtt{C4}$ vs. pitch $\mathtt{E4}$ following $\mathtt{C4}$. The way to accomplish this is to have the following term in the objective function
\begin{equation}\label{eq:objective}
    -\sum_{ \substack{i\in [n-1]  \\ j,j' \in P}} W_{j,j'} x_{i,j} x_{i+1,j'}.
\end{equation}
Here, $W_{j,j'}$ is the weight associated with having note $j'$ after note $j$. The larger the weight, the higher the reward we have in the objective function. The weights can be determined by analyzing some music pieces and forming a transition matrix of weights examining the consecutive notes. Below, we illustrate this idea through a simple music piece.

\begin{figure}[h]
	\begin{center}
		\includegraphics[scale=0.6]{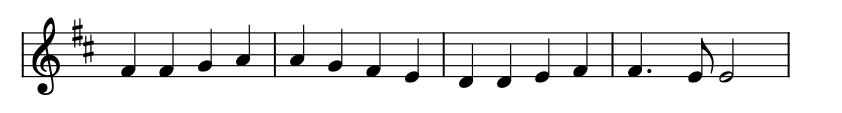}
		\caption{An excerpt from Beethoven's Ode to Joy.}
		\label{fig:odetojoy}
	\end{center}
\end{figure}

In Figure \ref{fig:odetojoy}, an excerpt from Beethoven's Ode to Joy is given. To identify the weights, we count the occurrences of consecutive pairs. To start with, we identify the occurrences of each note and then count the number of times the note is followed by another particular note. For instance, the note $\mathtt{F\#4}$ appears at positions 1,2,7,12,13. It is followed by  $\mathtt{F\#4}$ and $\mathtt{E4}$ twice, and by $\mathtt{G4}$ once. Hence, we can deduce the following weights:
\begin{equation}\label{eq:weights}
W_{F\#4,E4} = 2, ~~W_{F\#4,F\#4} = 2,~~W_{F\#4,G4} = 1.
\end{equation}
As the selected music piece is a short one, it does not contain all the notes from $P$ and for those notes $i$, the corresponding weight $W_{i,j}=0$ for all $j$. We would like to remark that having 0 as the weight does not imply that the corresponding note tuple is always avoided but means that we don't give additional rewards to such pairs. The list of all non-zero weights are as follows:
\begin{align}\label{eq:probabilities}
&M_{F\#4,E4} = 2, ~~W_{F\#4,F\#4} = 2,~~W_{F\#4,G4} = 1,\\
&W_{G4,F\#4}  = 1,~~ W_{G4,A4}= 1, \nonumber\\
&W_{A4,G4} = 1,~~~~W_{A4,A4} = 1, \nonumber\\
&W_{E4,D4} = 1,~~~~W_{E4,E4} =1 ,~~~~~W_{E4,F\#4} = 1,\nonumber\\
&W_{D4,D4} = 1,~~~~W_{D4,E4} = 1\nonumber.
\end{align}
Note that taking the number of occurrences as the weights, we are also giving rewards to pitches that appear more frequently than the other. For instance, $\mathtt{F\#}$ appears five times, and the overall reward is higher when more $\mathtt{F\#}$'s appear in the sequence.

Now the question is how should we select set $P$. We can use Eq.~\ref{eq:modeldiatonic} to define our binary variables as the degrees from a scale. Hence, the matrix $W$ defines the transition weights between the scale degrees. Note that the newly generated music piece mimics the one from which the transition weights are obtained. The longer the music piece, the better estimates are obtained for the weights. Multiple pieces can be selected as well, paying attention that they are from the same scale, or in case considering pieces from different scales, one should take degrees of the scale instead of the pitches when calculating the weights.

\begin{figure}[h]
	\begin{center}
		\includegraphics[scale=0.6]{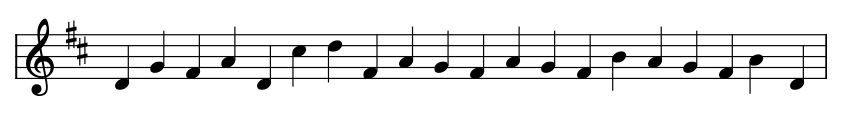}
		\caption{The degree sequence is obtained from D-Wave using the transition weights from Ode to Joy and it is interpreted in $\mathtt{D~Major}$. }
		\label{fig:odetojoy_prob}
	\end{center}
\end{figure}

We define the QUBO formulation defined through binary variables given in Eq.~\ref{eq:modeldiatonic} using the constraints defined in Eq.~\ref{eq:single_note}, Eq.~\ref{eq:firstlast}, Eq.~\ref{eq:samenote}, Eq.~\ref{eq:tendency} and  Eq.~\ref{eq:objective} as the objective function with the weights obtained from Eq.~\ref{eq:probabilities}. Note that in this case, one needs to properly set the penalty coefficients; in case the constraint is violated and there is a reward, an increase in the energy due to the penalty should be larger than the decrease in the energy due to reward. The resulting music piece obtained from the D-Wave hybrid solver is given in Figure \ref{fig:odetojoy_prob}. We interpreted the obtained degree sequence in $\mathtt{D~Major}$.

Alternatively, instead of generating weights for the transitions between individual degrees, we can collect statistics about the different intervals used in the music piece and how often they appear. Then accordingly, we can reward the intervals that occur more frequently.

\subsubsection{Rests}

The sets we have considered so far only consisted of the pitches; however, we may want to include rests in the music piece as well. In this case, a rest element can be included in the set $P$ with appropriate rules. For instance, we may want to avoid two consecutive rests, which we can easily accomplish with the rules we have shown for consecutive notes. In addition, we can limit the number of rests used in the music piece by introducing the following constraint:
\begin{equation}
    \sum_{i} x_{i,r} = k,
\end{equation}
where $k$ is the total number of rests we want in the music piece and $x_{i,r}$ denotes that note at position $i$ is a rest. 

\subsection{Rhythm Generation}

So far, we have discussed generating the pitches of the melody. In this section, we additionally take into account the rhythm.

\subsubsection{Rhythmic Sequence}
When generating a music pieace, one option would be to generate the pitch sequence and the rhythmic sequence separately. In such a case, the idea will be similar to what we had previously. The set $S$ will consists of possible durations for the notes, such as whole, half, quarter etc. The binary variables $y_{i,j}$ will be defined as 
\begin{equation}\label{eq:modelrhythm}
y_{i,j} =  \begin{cases}%
1      & \text{note at position $i$ has duration $j$,}\\
0 & \text{otherwise.}
\end{cases}
\end{equation}
for $i \in [n]$ and $j \in D$. 
 
Similarly, the first rule we need to incorporate is that only one of the variables $y_{i,j}$ is equal to 1 for each position $i$, which is expressed using the following constraint for each $i \in [n]$:

\begin{equation}\label{eq:singlerhythm}
    \sum_{j \in D} y_{i,j} = 1.
\end{equation}

For the objective function, the same method can be used. Let us denote half note by $\mathtt{H}$, quarter note by $\mathtt{Q}$, dotted quarter note by $\mathtt{DQ}$ and eighth note by $\mathtt{E}$. For the music piece given in Figure \ref{fig:odetojoy}, we obtain the following weights:
\begin{align}\label{eq:probsd}
    &W_{Q,Q} = 11,~~W_{Q,DQ} = 1 \\
    &W_{DQ,E} = 1 \\
    &W_{E,H} = 1.
\end{align}
If we only incorporate Eq.~\ref{eq:singlerhythm} and Eq.~\ref{eq:probsd} in our formulation, then it is very likely that we will have a sequence of quarter notes only, as they have the highest weight. To avoid this, we will make sure that there are at least two notes of each duration using the following constraint for each $d \in D$:
\begin{equation}\label{eq:two}
    \sum_{i=1}^n y_{i,d} \geq 2.
\end{equation}

\begin{figure}[h]
\begin{center}
\includegraphics[scale=0.6]{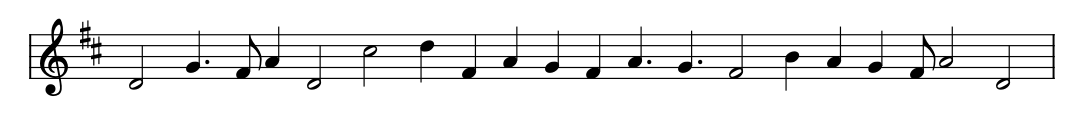}
\caption{The rhythmic sequence obtained from D-Wave using is combined with the pitch sequence obtained in Fig. \ref{fig:odetojoy_prob}.}
\label{fig:odetojoy_rhyhthm}
\end{center}
\end{figure}

We generate a rhythmic sequence with the binary variables defines as in Eq.~\ref{eq:modelrhythm}, using Eq.~\ref{eq:singlerhythm} and Eq.~\ref{eq:two} as the constraints and the transition weights given in Eq.~\ref{eq:probsd} to obtain the objective function defined in Eq.~\ref{eq:objective}. We combine it with the degree sequence generated in Fig. \ref{fig:odetojoy_prob} and obtain the music piece given in Fig. \ref{fig:odetojoy_rhyhthm}.

We would like to note that in this approach, we are not fixing the total length of the music piece, but we are determining the durations for a given fixed set of notes. To have a fixed music length, what we can do is include a constraint that takes into account the duration of each type of note:
\begin{equation}
    \sum_{\substack{i \in [n] \\ j \in D}} d(j)y_{i,j} = L,
\end{equation}
where $d(j)$ is the duration of $j$ and $L$ is the total length of the music piece in terms of eighth notes so that $d(\mathtt{E})=1$, $d(\mathtt{Q})=2$, $d(\mathtt{DQ})=3$ and $d(\mathtt{H})=4$. We discretize the durations in terms of eighth notes as it is the note with shortest duration in our example.

\subsubsection{Rhythm and Pitch Generated Together}

Another alternative is to consider pitches together with their durations. If $P$ is the set of possible pitches, and $D$ is the set of possible durations, then overall, there will be $|D||P|$ possibilities for each note. Assuming there are $n$ notes, the number of variables we need significantly increases to $n|D||P|$ in this case. The binary variables take the following form:
\begin{equation}
x_{i,j,k} =  \begin{cases}%
1      & \text{note at position $i$ is pitch $j$ and has duration $k$,}\\
0 & \text{otherwise.}
\end{cases}
\end{equation}
for $i \in [n]$, $j \in P$ and $k \in D$. 

The previous rules we defined about the consecutive notes and intervals apply here too. Those constraints are included independent of $k$, the note's duration. 

Likewise, we can still take an objective function based on another music piece. This time, we inspect the number of occurrences of consecutive pitch-duration pairs.

\subsection{Harmony Generation}

We have seen so far how to generate melody and rhythm. Now it is time to discuss harmonization using quantum annealing.

\input{markovmusic}

\subsubsection{qHarmony}

One of the few existing pieces of literature on music composition using quantum annealing is provided by Kirke et al. \cite{kirke2017experiments}. Taking a music piece from $\mathtt{C~Major}$, the qHarmony tool harmonizes the given melody. In this section, we will briefly describe the used technique.

Instead of QUBO formulation, the problem is encoded directly using Ising formulation, taking also into account the embedding to the Chimera graph, the graph representing the topology of the underlying qubits of the D-Wave 2000 QPU. For simplicity, we will not discuss the embedding component as this part can be handled by D-Wave solvers. The set $S$ is defined as $\{\mathtt{C4},\mathtt{D4},\mathtt{E4}, \mathtt{F4}, \mathtt{G4}, \mathtt{A4}, \mathtt{B4}, \mathtt{C5}\}$.\footnote{In the original text, the set is taken as $\{c,d,e,f,g,a,b,C\}$ where $C$ denotes the note one octave above $c$.} 

Unlike the approach we have followed so far where we were considering a sequence of $n$ notes at a time, only a single chord is generated at a time. The spin variables $q_i$ for $i \in S$ are defined as in Eq.~\ref{eq:spin}.
\begin{equation}\label{eq:spin}
q_{i} =  \begin{cases}%
1     & \text{note $i$ is selected,}\\
-1 & \text{otherwise.}
\end{cases}
\end{equation}

The coupling constants are defined as $J_{i,j} = 7-2|i-j|$. Note that the smaller the $|i-j|$ is, the larger the value of $J_{i,j}$. Since the annealer tries to minimize the overall energy, the qubits corresponding to the notes that are semitone or tone apart will have a higher coupling constant; thus, the possibility of getting such intervals will reduce. 

The external magnetic field $h_i$'s indicate which note is harmonized; it is set as $h_i=-7$, if note $i$ is harmonized, and 1, otherwise. Hence setting $q_i$ as 1 will reduce the energy as expected. Note $i$ will be called the input note. 

It is also possible to specify multiple input notes to harmonize by setting more than one $h_i$ value as $-7$. Furthermore, the tool works in a hybrid manner involving some classical components as well, allowing the user to identify for which notes of the piece, harmony should be generated, and whether it should be a minor or major chord. 

There are several drawbacks of the proposed approach. First of all, the provided example works for eight notes, in particular the notes from the scale of $\mathtt{C~Major}$. This is not a significant problem as the given example can be generalized for other sets and scales by tuning the coupling constants. Secondly, the generated chords at each time point do not take into account the previously generated ones. Finally, the number of notes that will appear in the chord can not be determined beforehand in this model. Next, we will discuss an alternative approach and try to overcome the mentioned drawbacks.
\subsubsection{Model}

Suppose that we would like to add three-note chords, triads, to a given music piece. The notes appearing in the triads will be restricted to the set of pitches $P=\{1,2,3,4,5,6,7,8\}$, where the elements represent the degrees of the scale. Our binary variables $x_{i,j}$ for $i \in [n]$ and $j \in P$ are defined as
\begin{equation}\label{eq:model_chords}
x_{i,j} =  \begin{cases}%
1      & \text{chord at position $i$ contains note $j$,}\\
0 & \text{otherwise.}
\end{cases}
\end{equation}

At each time point, we would like to have three notes. Hence, we are going to add the following constraint to our model:
\begin{equation}\label{eq:three_notes}
    \sum_{j \in P} x_{i,j} = 3.
\end{equation}

Recall that we had already defined a constraint so that the melody begins and ends with the first degree of the scale. Similarly, we can enforce the first and the last triads to be built upon the first degree of the scale.  We can add the terms defined in Eq.~\ref{eq:first_last_chord} to the objective function:
\begin{equation}\label{eq:first_last_chord}
(1-x_{1,1}),~~(1-x_{1,3}),~~(1-x_{1,5}),~~(1-x_{n,1}),~~(1-x_{n,3}),~~(1-x_{n,5}).
\end{equation}

For each time point, one of the pitches in the triad should be the pitch of the melody. Let $N=[p_{1_j},p_{2_j},\dots,p_{n_j} ]$ be the note sequence that we would like to harmonize. We assume that each $p_{i_j} \in P $. To enforce that one of the three notes in the triad at time point $i$ is $p_{i_j}$, we add the term defined in Eq.~\ref{eq:note_i} to the objective function for all $i= 2,\dots,n-1$. (We are assuming that the first and the last pitch of the note sequence is the first degree of the scale.)
\begin{equation}\label{eq:note_i}
(1-x_{i,p_{i_j}})
\end{equation}

We want to generate triads either obtained by taking the degrees of the scale as their root or their inversions. We want to avoid consecutive degrees from the scale and the degrees 1-7, 2-8, 1-8 to appear in the same triad by adding the terms defined in Eq.~\ref{eq:triad} to the objective function 
for all $i= 2,\dots,n-1$, $|j-j'| \in \{1,6,7\}$.
\begin{equation}\label{eq:triad}
x_{i,j}x_{i,j'}
\end{equation}

\begin{figure}[h]
\begin{center}
\includegraphics[scale=0.5]{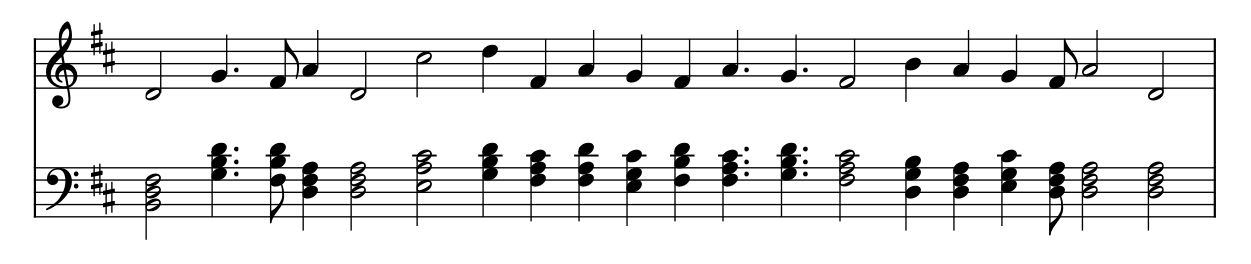}
\caption{The music piece obtained by harmonizing the music piece given in Fig. \ref{fig:odetojoy_rhyhthm}}
\label{fig:harmony}
\end{center}
\end{figure}

Using \cref{eq:model_chords,eq:three_notes,eq:first_last_chord,eq:note_i,eq:triad} and harmonizing the music piece illustrated in Fig. \ref{fig:odetojoy_rhyhthm}, we obtain the music piece presented in Fig. \ref{fig:harmony}.

The new method allows more flexibility than the one previously proposed. Yet, it does not take into account any preference between the consecutive chords. For instance, it is likely that some chords built upon the particular degrees of a scale follow each other, and sometimes it is the opposite case. Furthermore, if one follows 4-part harmony rules, there should be further rules that should be incorporated in the model.

%% file: markovmusic.tex
\subsubsection{Algorhythms}

Although it does not directly fit into the scope of harmonization, we would like to discuss a previous work that uses Markov Random Fields to produce chord progressions. ``Algorhythms: Generating Music with D-Wave's Quantum Annealer" project was submitted to the iQuHack 2021 MIT Hackathon \cite{hackaton}, and it features a novel example of how one can use a Markov Network in order to generate music. 

\begin{figure}[h!]
\begin{center}
\includegraphics[scale=0.5]{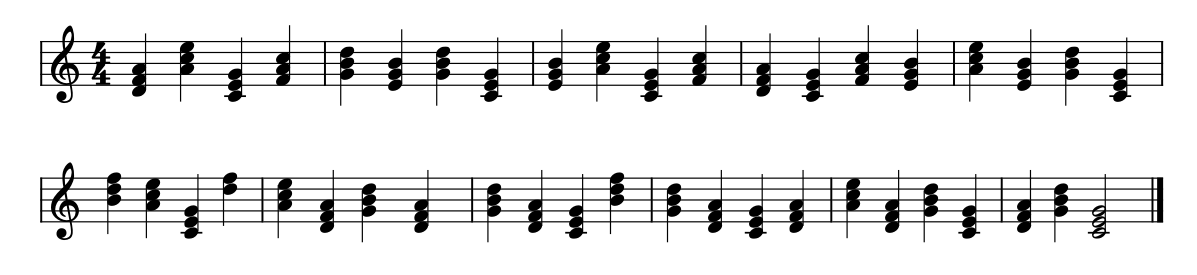}
\caption{A song containing a sequence of chords, submitted to the iQuHack 2021 MIT Hackathon by the authors of the ``Algorhythms: Generating Music with D-Wave's Quantum Annealer" project. \cite{hackaton}}
\label{fig:song}
\end{center}
\end{figure}

Let us explain the idea briefly. Each node represents one of the chords $\mathtt{I}, \mathtt{II}, \mathtt{III}, \mathtt{IV}, \mathtt{V}, \mathtt{VI}, \mathtt{VIIdim}$. Having realized that chords can represent the states of a Markov Random Field, the authors created a mechanism that assigned values to the potentials between the nodes in a random manner. The potentials for $(0,0)$ and $(1,1)$ are assigned high values to avoid two chords or no chord being selected at a time point. Each run of the algorithm outputs a chord based on the randomly assigned potentials. To get a sequence of chords, the algorithm is called several times. In Fig. \ref{fig:song}, we can see a song generated with this algorithm.

Each chord is selected from the set of samples that the D-Wave computer outputs. What we are observing is a sequence of states each of which represents the lowest energy state for the given configuration of the potentials. As each chord is generated independently, the progressions do not follow any rules or music style. Next we will see how we can create a chord progression following some rules.

,

\subsubsection{Generating a Chord Sequence Using Markov Random Fields}

Consider the chords that can be built taking each note of the major scale as the root of a triad. To generate a chord progression of $n$ notes we need to define our nodes as $\mathtt{I}_i, \mathtt{ii}_i, \mathtt{iii}_i, \mathtt{IV}_i, \mathtt{V}_i, \mathtt{vi}_i, \mathtt{VIIdim}_i$ where $i\in [n]$. For each time point, we have multiple nodes reflecting multiple possibilities. Then, we can employ a more structured chord progression defining the potentials accordingly. 

One way to make the result more similar to the musical pieces we are used to listening to is to take into account some rules of musical harmony. For example, there are musical structures that have existed for a long time and remain in force due to the stability they provide when presenting a complete musical idea: the cadences. Harmonic \textit{cadences} are progressions of two or more chords that help complete a musical section by giving resolution to a musical phrase. We have, for instance, the ``perfect" cadence formed by the chords $\mathtt{V}-\mathtt{I}$. This structure is one of the most used because of its forcefulness, but there are also other cadences that help complete a musical discourse. If we were to encode these structures into our network, we would need to set low energy for the potentials relating, in the case of the perfect cadence, the chords $\mathtt{V}_{i-1}$ and $\mathtt{I}_i$. This would ensure that when the system evolves, it would end up in a state that contains this particular progression.

For instance, suppose that we want to generate a sequence of four chords that will tend to exhibit the pattern of a perfect cadence. We can start by identifying our binary random variables and defining the couplings between them. Note that in order to avoid having several chords played at the same time, we have to set high values for the potentials that relate to the chords designated for each step of the sequence, i.e,  the set of chords that represent the first chord, the second chord, etc. Let's visualize the graph of our problem.

\begin{figure}[h!]
	\begin{center}
		\includegraphics[scale=0.3]{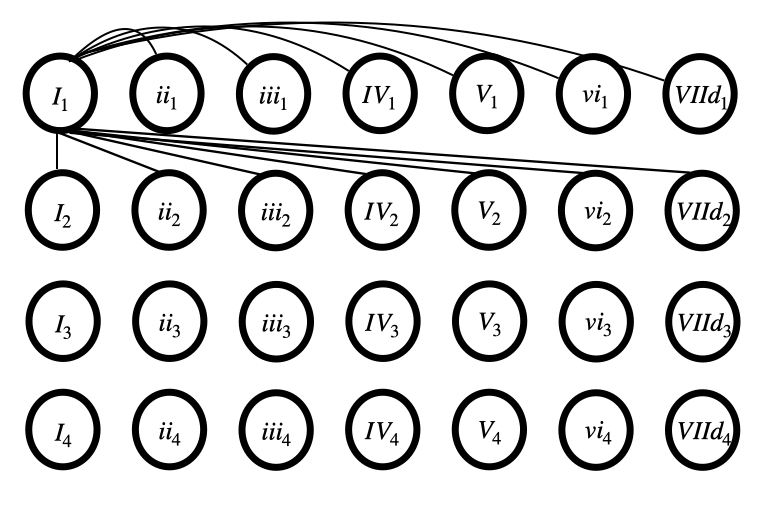}
		\caption{Markov Random Field is delineated with the random variables defined for a progression of 4 chords. For simplicity, the dependency relations are shown only for the chord $\mathtt{I}_1$ (1'st chord, chord $\mathtt{I}$). }
		\label{fig:augmented}
	\end{center}
\end{figure}

Now, let's define our pairwise potentials. We give some examples in Tables~\ref{table:p1}, \ref{table:p2} and \ref{table:p3}. Table \ref{table:p1}, observe that we penalize the configuration that represents having both chords played at the same time. In \ref{table:p2},  we penalize the configuration that represents having the same chord repeated over in the sequence. In \ref{table:p2}, we favor the configuration that represents having the pattern $\mathtt{V}$-$\mathtt{I}$. The rest of the potentials are defined in a similar manner.

\begin{table}[h!]
\centering
\begin{tabular}{ |c|c|c| } 
\hline
 $\mathtt{I}_1$ & $\mathtt{ii}_1$ & p( $\mathtt{I}_1$, $\mathtt{ii}_i$) \\
\hline
0 & 0 & 50.0 \\ 
\hline
0 & 0 & 50.0 \\
\hline
1 & 0 & 50.0 \\
\hline
1 & 1 & 100.0 \\
\hline
\end{tabular}
\caption{Values of the potential that associates the chords $\mathtt{I}_1$ and $\mathtt{ii}_1$.}
\label{table:p1}
\end{table}

\begin{table}[h!]
\centering
\begin{tabular}{ |c|c|c| } 
\hline
 $\mathtt{I}_1$ & $\mathtt{I}_2$ & $p( \mathtt{I}_1, \mathtt{I}_2)$ \\
\hline
0 & 0 & 50.0 \\ 
\hline
0 & 0 & 50.0 \\
\hline
1 & 0 & 50.0 \\
\hline
1 & 1 & 100.0 \\
\hline
\end{tabular}
\caption{Values of the potential that associates the chords $\mathtt{I}_1$ and $\mathtt{I}_2$.} 
\label{table:p2}
\end{table}

\begin{table}[h!]
\centering
\begin{tabular}{ |c|c|c| } 
\hline
$\mathtt{V}_1$ & $\mathtt{I}_2$  & $p(\mathtt{V}_1,\mathtt{I}_2)$ \\
\hline
0 & 0 & 50.0 \\ 
\hline
0 & 0 & 50.0 \\
\hline
1 & 0 & 50.0 \\
\hline
1 & 1 & 0.0 \\
\hline
\end{tabular}
\caption{Values of the potential that associates the chords $\mathtt{V}_1$ and $\mathtt{I}_2$ }
\label{table:p3}
\end{table}

\begin{figure}[h!]
	\begin{center}
		\includegraphics[scale=0.6]{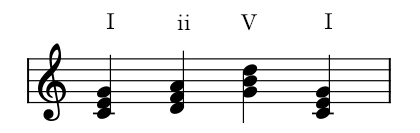}
		\caption{A chord progression generated using Markov Random Fields incorporating the perfect cadence.}
		\label{fig:song2}
	\end{center}
\end{figure}

This approach is rather different from the one used in the ``Algorhythms: Generating Music with D-Wave's Quantum Annealer" project. First, we do not have to run the algorithm several times because we are generating the sequence all at once. Then, observe that we are setting specific values to the potentials and that by doing so, we are encoding some harmony considerations into the model. One of the results we can get with this algorithm is given in Fig.~\ref{fig:song2}. Observe that the resulting progression satisfies the requirement of showing the pattern $\mathtt{V}$-$\mathtt{I}$.

In conclusion, as we have seen, Markov Chains and Markov Random Fields are tools that prove to be useful when solving optimization problems. Moreover, they also represent new approaches to the task of music generation.


%% file: discussion.tex
The main goal of this chapter was to lay the groundwork for music composition using quantum annealing. We have introduced various formulations for melody, rhythm, and harmony generation and demonstrated how one might incorporate any music rule into the model. The presented methodologies will be of interest to composers, quantum computing researchers, and the community of algorithmic music generation and allow interested readers to build upon the given notions.  As the first comprehensive study focusing on quantum annealing, we believe that it will foster the development of the field. As a side contribution, we demonstrate a new application area for quantum annealing, where creativity is in the foreground instead of potential quantum speedup. 

The study has opened up a vast amount of directions to investigate.  First of all, the models can be enhanced to incorporate further rules. This might be beneficial for harmony generation in particular. Furthermore, one can consider the approach we have taken for defining the objection function while generating a melody for the Markov Random Fields. From an existing music piece, the harmonic progressions can be identified, and the potentials can be defined accordingly. One may focus on generating counterpoint music, which heavily depends on rules. As a natural progression, a target problem would be music completion. Music completion refers to the task of replacing the missing notes in a given music piece. It is a suitable candidate problem as it can be defined as the problem of maximizing the likelihood that a sequence of notes replaces the missing ones. Besides music creation, one may consider other problems from the domain of music, such as music arrangement, where the aim is to arrange a given music piece to a given list of instruments by reduction. 

%% file: acknowledgement.tex
ÖS and LB have been partially supported by Polish National Science Center under the grant agreement 2019/33/B/ST6/02011. 

The project was initiated under the QIntern program organized by QWorld, therefore we would like to thank the organizers of the program.

We would like to thank Adam Glos and  Jarosław  Adam Miszczak for their valuable comments.